\documentclass[12pt]{article}

\makeatletter
\newcount\@hour\newcount\@minute\chardef\@x10\chardef\@xv60
\def\tcitime{
\def\@time{%
  \@minute\time\@hour\@minute\divide\@hour\@xv
  \ifnum\@hour<\@x 0\fi\the\@hour:%
  \multiply\@hour\@xv\advance\@minute-\@hour
  \ifnum\@minute<\@x 0\fi\the\@minute
  }}%
\def\QCTOpt[#1]#2{%
  \def\QCTOptB{#1}
  \def\QCTOptA{#2}
}
\def\QCTNOpt#1{%
  \def\QCTOptA{#1}
  \let\QCTOptB\empty
}
\def\Qct{%
  \@ifnextchar[{%
    \QCTOpt}{\QCTNOpt}
}
\def\QCBOpt[#1]#2{%
  \def\QCBOptB{#1}
  \def\QCBOptA{#2}
}
\def\QCBNOpt#1{%
  \def\QCBOptA{#1}
  \let\QCBOptB\empty
}
\def\Qcb{%
  \@ifnextchar[{%
    \QCBOpt}{\QCBNOpt}
}
\def\PrepCapArgs{%
  \ifx\QCBOptA\empty
    \ifx\QCTOptA\empty
      {}%
    \else
      \ifx\QCTOptB\empty
        {\QCTOptA}%
      \else
        [\QCTOptB]{\QCTOptA}%
      \fi
    \fi
  \else
    \ifx\QCBOptA\empty
      {}%
    \else
      \ifx\QCBOptB\empty
        {\QCBOptA}%
      \else
        [\QCBOptB]{\QCBOptA}%
      \fi
    \fi
  \fi
}
\newcount\GRAPHICSTYPE
\GRAPHICSTYPE=\z@
\def\GRAPHICSPS#1{%
 \ifcase\GRAPHICSTYPE
   \special{ps: #1}%
 \or
   \special{language "PS", include "#1"}%
 \fi
}%
\def\graffile#1#2#3#4{%
    \leavevmode
    \raise -#4 \BOXTHEFRAME{%
        \hbox to #2{\raise #3\hbox{\null #1}}}%
}%
\def\draftbox#1#2#3#4{%
 \leavevmode\raise -#4 \hbox{%
  \frame{\rlap{\protect\tiny #1}\hbox to #2%
   {\vrule height#3 width\z@ depth\z@\hfil}%
  }%
 }%
}%
\newcount\draft
\draft=\z@

\newif\ifwasdraft
\wasdraftfalse

\def\GRAPHIC#1#2#3#4#5{%
 \ifnum\draft=\@ne\draftbox{#2}{#3}{#4}{#5}%
  \else\graffile{#1}{#3}{#4}{#5}%
  \fi
 }%
\def\addtoLaTeXparams#1{%
    \edef\LaTeXparams{\LaTeXparams #1}}%
\newif\ifBoxFrame \BoxFramefalse
\newif\ifOverFrame \OverFramefalse
\newif\ifUnderFrame \UnderFramefalse

\def\BOXTHEFRAME#1{%
   \hbox{%
      \ifBoxFrame
         \frame{#1}%
      \else
         {#1}%
      \fi
   }%
}

\def\doFRAMEparams#1{\BoxFramefalse\OverFramefalse\UnderFramefalse\readFRAMEparams#1\end}%
\def\readFRAMEparams#1{%
 \ifx#1\end%
  \let\next=\relax
  \else
  \ifx#1i\dispkind=\z@\fi
  \ifx#1d\dispkind=\@ne\fi
  \ifx#1f\dispkind=\tw@\fi
  \ifx#1t\addtoLaTeXparams{t}\fi
  \ifx#1b\addtoLaTeXparams{b}\fi
  \ifx#1p\addtoLaTeXparams{p}\fi
  \ifx#1h\addtoLaTeXparams{h}\fi
  \ifx#1X\BoxFrametrue\fi
  \ifx#1O\OverFrametrue\fi
  \ifx#1U\UnderFrametrue\fi
  \ifx#1w
    \ifnum\draft=1\wasdrafttrue\else\wasdraftfalse\fi
    \draft=\@ne
  \fi
  \let\next=\readFRAMEparams
  \fi
 \next
 }%
\def\IFRAME#1#2#3#4#5#6{%
      \bgroup
      \let\QCTOptA\empty
      \let\QCTOptB\empty
      \let\QCBOptA\empty
      \let\QCBOptB\empty
      #6%
      \parindent=0pt%
      \leftskip=0pt
      \rightskip=0pt
      \setbox0 = \hbox{\QCBOptA}%
      \@tempdima = #1\relax
      \ifOverFrame
          \typeout{This is not implemented yet}%
          \show\HELP
      \else
         \ifdim\wd0>\@tempdima
            \advance\@tempdima by \@tempdima
            \ifdim\wd0 >\@tempdima
               \textwidth=\@tempdima
               \setbox1 =\vbox{%
                  \noindent\hbox to \@tempdima{\hfill\GRAPHIC{#5}{#4}{#1}{#2}{#3}\hfill}\\%
                  \noindent\hbox to \@tempdima{\parbox[b]{\@tempdima}{\QCBOptA}}%
               }%
               \wd1=\@tempdima
            \else
               \textwidth=\wd0
               \setbox1 =\vbox{%
                 \noindent\hbox to \wd0{\hfill\GRAPHIC{#5}{#4}{#1}{#2}{#3}\hfill}\\%
                 \noindent\hbox{\QCBOptA}%
               }%
               \wd1=\wd0
            \fi
         \else
            \ifdim\wd0>0pt
              \hsize=\@tempdima
              \setbox1 =\vbox{%
                \unskip\GRAPHIC{#5}{#4}{#1}{#2}{0pt}%
                \break
                \unskip\hbox to \@tempdima{\hfill \QCBOptA\hfill}%
              }%
              \wd1=\@tempdima
           \else
              \hsize=\@tempdima
              \setbox1 =\vbox{%
                \unskip\GRAPHIC{#5}{#4}{#1}{#2}{0pt}%
              }%
              \wd1=\@tempdima
           \fi
         \fi
         \@tempdimb=\ht1
         \advance\@tempdimb by \dp1
         \advance\@tempdimb by -#2%
         \advance\@tempdimb by #3%
         \leavevmode
         \raise -\@tempdimb \hbox{\box1}%
      \fi
      \egroup%
}%
\def\DFRAME#1#2#3#4#5{%
 \begin{center}
     \let\QCTOptA\empty
     \let\QCTOptB\empty
     \let\QCBOptA\empty
     \let\QCBOptB\empty
     \ifOverFrame 
        #5\QCTOptA\par
     \fi
     \GRAPHIC{#4}{#3}{#1}{#2}{\z@}
     \ifUnderFrame 
        \par #5\QCBOptA
     \fi
 \end{center}%
 }%
\def\FFRAME#1#2#3#4#5#6#7{%
 \begin{figure}[#1]%
  \let\QCTOptA\empty
  \let\QCTOptB\empty
  \let\QCBOptA\empty
  \let\QCBOptB\empty
  \ifOverFrame
    #4
    \ifx\QCTOptA\empty
    \else
      \ifx\QCTOptB\empty
        \caption{\QCTOptA}%
      \else
        \caption[\QCTOptB]{\QCTOptA}%
      \fi
    \fi
    \ifUnderFrame\else
      \label{#5}%
    \fi
  \else
    \UnderFrametrue%
  \fi
  \begin{center}\GRAPHIC{#7}{#6}{#2}{#3}{\z@}\end{center}%
  \ifUnderFrame
    #4
    \ifx\QCBOptA\empty
      \caption{}%
    \else
      \ifx\QCBOptB\empty
        \caption{\QCBOptA}%
      \else
        \caption[\QCBOptB]{\QCBOptA}%
      \fi
    \fi
    \label{#5}%
  \fi
  \end{figure}%
 }%
\newcount\dispkind%
\def\FRAME#1#2#3#4#5#6#7#8{%
 \ifnum\draft=\@ne
   \wasdrafttrue
 \else
   \wasdraftfalse%
 \fi
 \def\LaTeXparams{}%
 \dispkind=\z@
 \def\LaTeXparams{}%
 \doFRAMEparams{#1}%
 \ifnum\dispkind=\z@\IFRAME{#2}{#3}{#4}{#7}{#8}{#5}\else
  \ifnum\dispkind=\@ne\DFRAME{#2}{#3}{#7}{#8}{#5}\else
   \ifnum\dispkind=\tw@
    \edef\@tempa{\noexpand\FFRAME{\LaTeXparams}}%
    \@tempa{#2}{#3}{#5}{#6}{#7}{#8}%
    \fi
   \fi
  \fi
  \ifwasdraft\draft=1\else\draft=0\fi{}%
 }%
\def\TEXUX#1{"texux"}

\long\def\QQQ#1#2{%
     \long\expandafter\def\csname#1\endcsname{#2}}%
\@ifundefined{QTP}{\def\QTP#1{}}{}
\@ifundefined{Qlb}{}{}
\@ifundefined{Qlt}{}{}
\long\def\QQA#1#2{}%
\def\QTR#1#2{{\csname#1\endcsname #2}}
\long\def\TeXButton#1#2{#2}%
\def\EXPAND#1[#2]#3{}%
\def\NOEXPAND#1[#2]#3{}%
\def\LaTeXparent#1{}%
\def\ChildStyles#1{}%
\def\ChildDefaults#1{}%
\def\QTagDef#1#2#3{}%
\def\QQfnmark#1{\footnotemark}

\@ifundefined{INDEX}{\def\INDEX#1#2{}{}}{}%
\@ifundefined{SUBINDEX}{\def\SUBINDEX#1#2#3{}{}{}}{}%
\def\initial#1{\bigbreak{\raggedright\large\bf #1}\kern 2\p@
   \penalty3000}%
\@ifundefined{ZZZ}{}{\makeatletter\input gnuindex.sty\makeatother\makeindex\makeatletter}%
\@ifundefined{abstract}{%
 \def\abstract{%
  \if@twocolumn
   \section*{Abstract (Not appropriate in this style!)}%
   \else \small 
   \begin{center}{\bf Abstract\vspace{-.5em}\vspace{\z@}}\end{center}%
   \quotation 
   \fi
  }%
 }{%
 }%
\@ifundefined{endabstract}{\def\endabstract
  {\if@twocolumn\else\endquotation\fi}}{}%
\@ifundefined{maketitle}{\def\maketitle#1{}}{}%
\@ifundefined{affiliation}{\def\affiliation#1{}}{}%
\@ifundefined{proof}{}{}%
\@ifundefined{endproof}{}{}%
\@ifundefined{newfield}{\def\newfield#1#2{}}{}%
\@ifundefined{chapter}{\def\chapter#1{\par(Chapter head:)#1\par }%
 \newcount\c@chapter}{}%
\@ifundefined{part}{\def\part#1{\par(Part head:)#1\par }}{}%
\@ifundefined{section}{\def\section#1{\par(Section head:)#1\par }}{}%
\@ifundefined{subsection}{\def\subsection#1%
 {\par(Subsection head:)#1\par }}{}%
\@ifundefined{subsubsection}{\def\subsubsection#1%
 {\par(Subsubsection head:)#1\par }}{}%
\@ifundefined{paragraph}{\def\paragraph#1%
 {\par(Subsubsubsection head:)#1\par }}{}%
\@ifundefined{subparagraph}{\def\subparagraph#1%
 {\par(Subsubsubsubsection head:)#1\par }}{}%
\@ifundefined{therefore}{}{}%
\@ifundefined{backepsilon}{}{}%
\@ifundefined{yen}{}{}%
\@ifundefined{registered}{%
   \def\registered{\relax\ifmmode{}\r@gistered
                    \else$\m@th\r@gistered$\fi}%
 \def\r@gistered{^{\ooalign
  {\hfil\raise.07ex\hbox{$\scriptstyle\rm\text{R}$}\hfil\crcr
  \mathhexbox20D}}}}{}%
\@ifundefined{Eth}{}{}%
\@ifundefined{eth}{}{}%
\@ifundefined{Thorn}{}{}%
\@ifundefined{thorn}{}{}%
\@ifundefined{degree}{}{}%
\newdimen\theight
\def\Column{%
 \vadjust{\setbox\z@=\hbox{\scriptsize\quad\quad tcol}%
  \theight=\ht\z@\advance\theight by \dp\z@\advance\theight by \lineskip
  \kern -\theight \vbox to \theight{%
   \rightline{\rlap{\box\z@}}%
   \vss
   }%
  }%
 }%
\def\qed{%
 \ifhmode\unskip\nobreak\fi\ifmmode\ifinner\else\hskip5\p@\fi\fi
 \hbox{\hskip5\p@\vrule width4\p@ height6\p@ depth1.5\p@\hskip\p@}%
 }%
\def\miss{\hbox{\vrule height2\p@ width 2\p@ depth\z@}}%
\def\tcol#1{{\baselineskip=6\p@ \vcenter{#1}} \Column}  %
\def\newfmtname{LaTeX2e}
\def\chkcompat{%
   \if@compatibility
   \else
     \usepackage{latexsym}
   \fi
}

\ifx\fmtname\newfmtname
  \DeclareOldFontCommand{\rm}{\normalfont\rmfamily}{\mathrm}
  \DeclareOldFontCommand{\sf}{\normalfont\sffamily}{\mathsf}
  \DeclareOldFontCommand{\tt}{\normalfont\ttfamily}{\mathtt}
  \DeclareOldFontCommand{\bf}{\normalfont\bfseries}{\mathbf}
  \DeclareOldFontCommand{\it}{\normalfont\itshape}{\mathit}
  \DeclareOldFontCommand{\sl}{\normalfont\slshape}{\@nomath\sl}
  \DeclareOldFontCommand{\sc}{\normalfont\scshape}{\@nomath\sc}
  \chkcompat
\fi

\def\alpha{\Greekmath 010B }%
\def\beta{\Greekmath 010C }%
\def\gamma{\Greekmath 010D }%
\def\delta{\Greekmath 010E }%
\def\epsilon{\Greekmath 010F }%
\def\zeta{\Greekmath 0110 }%
\def\eta{\Greekmath 0111 }%
\def\theta{\Greekmath 0112 }%
\def\iota{\Greekmath 0113 }%
\def\kappa{\Greekmath 0114 }%
\def\lambda{\Greekmath 0115 }%
\def\mu{\Greekmath 0116 }%
\def\nu{\Greekmath 0117 }%
\def\xi{\Greekmath 0118 }%
\def\pi{\Greekmath 0119 }%
\def\rho{\Greekmath 011A }%
\def\sigma{\Greekmath 011B }%
\def\tau{\Greekmath 011C }%
\def\upsilon{\Greekmath 011D }%
\def\phi{\Greekmath 011E }%
\def\chi{\Greekmath 011F }%
\def\psi{\Greekmath 0120 }%
\def\omega{\Greekmath 0121 }%
\def\varepsilon{\Greekmath 0122 }%
\def\vartheta{\Greekmath 0123 }%
\def\varpi{\Greekmath 0124 }%
\def\varrho{\Greekmath 0125 }%
\def\varsigma{\Greekmath 0126 }%
\def\varphi{\Greekmath 0127 }%

\def\nabla{\Greekmath 0272 }

\def\Greekmath#1#2#3#4{%
    \if@compatibility
        \ifnum\mathgroup=\symbold
           \mathchoice{\mbox{\boldmath$\displaystyle\mathchar"#1#2#3#4$}}%
                      {\mbox{\boldmath$\textstyle\mathchar"#1#2#3#4$}}%
                      {\mbox{\boldmath$\scriptstyle\mathchar"#1#2#3#4$}}%
                      {\mbox{\boldmath$\scriptscriptstyle\mathchar"#1#2#3#4$}}%
        \else
           \mathchar"#1#2#3#4%
        \fi 
    \else 
        \ifnum\mathgroup=5 
           \mathchoice{\mbox{\boldmath$\displaystyle\mathchar"#1#2#3#4$}}%
                      {\mbox{\boldmath$\textstyle\mathchar"#1#2#3#4$}}%
                      {\mbox{\boldmath$\scriptstyle\mathchar"#1#2#3#4$}}%
                      {\mbox{\boldmath$\scriptscriptstyle\mathchar"#1#2#3#4$}}%
        \else
           \mathchar"#1#2#3#4%
        \fi     	    
	  \fi}

\newif\ifGreekBold  \GreekBoldfalse
\let\SAVEPBF=\pbf
\def\pbf{\GreekBoldtrue\SAVEPBF}%

\@ifundefined{theorem}{}{}
\@ifundefined{lemma}{}{}
\@ifundefined{corollary}{}{}
\@ifundefined{conjecture}{}{}
\@ifundefined{proposition}{}{}
\@ifundefined{axiom}{}{}
\@ifundefined{remark}{}{}
\@ifundefined{example}{}{}
\@ifundefined{exercise}{}{}
\@ifundefined{definition}{}{}

\@ifundefined{mathletters}{%
  \newcounter{equationnumber}  
  \def\mathletters{%
     \addtocounter{equation}{1}
     \edef\@currentlabel{\theequation}%
     \setcounter{equationnumber}{\c@equation}
     \setcounter{equation}{0}%
     \edef\theequation{\@currentlabel\noexpand\alph{equation}}%
  }
  
}{}

\@ifundefined{BibTeX}{%
    \def\BibTeX{{\rm B\kern-.05em{\sc i\kern-.025em b}\kern-.08em
                 T\kern-.1667em\lower.7ex\hbox{E}\kern-.125emX}}}{}%
\@ifundefined{AmS}%
    {\def\AmS{{\protect\usefont{OMS}{cmsy}{m}{n}%
                A\kern-.1667em\lower.5ex\hbox{M}\kern-.125emS}}}{}%
\@ifundefined{AmSTeX}{}{}%
\ifx\ds@amstex\relax
\makeatother\endinput
\else
   \@ifpackageloaded{amstex}%
      {\message{amstex already loaded}\makeatother\endinput}
      {}
\fi
\let\DOTSI\relax
\def\RIfM@{\relax\ifmmode}%
\def\FN@{\futurelet\next}%
\newcount\intno@
\def\iint{\DOTSI\intno@\tw@\FN@\ints@}%
\def\iiint{\DOTSI\intno@\thr@@\FN@\ints@}%
\def\iiiint{\DOTSI\intno@4 \FN@\ints@}%
\def\idotsint{\DOTSI\intno@\z@\FN@\ints@}%
\def\ints@{\findlimits@\ints@@}%
\newif\iflimtoken@
\newif\iflimits@
\def\findlimits@{\limtoken@true\ifx\next\limits\limits@true
 \else\ifx\next\nolimits\limits@false\else
 \limtoken@false\ifx\ilimits@\nolimits\limits@false\else
 \ifinner\limits@false\else\limits@true\fi\fi\fi\fi}%
\def\multint@{\int\ifnum\intno@=\z@\intdots@                          
 \else\intkern@\fi                                                    
 \ifnum\intno@>\tw@\int\intkern@\fi                                   
 \ifnum\intno@>\thr@@\int\intkern@\fi                                 
 \int}
\def\multintlimits@{\intop\ifnum\intno@=\z@\intdots@\else\intkern@\fi
 \ifnum\intno@>\tw@\intop\intkern@\fi
 \ifnum\intno@>\thr@@\intop\intkern@\fi\intop}%
\def\intic@{%
    \mathchoice{\hskip.5em}{\hskip.4em}{\hskip.4em}{\hskip.4em}}%
\def\negintic@{\mathchoice
 {\hskip-.5em}{\hskip-.4em}{\hskip-.4em}{\hskip-.4em}}%
\def\ints@@{\iflimtoken@                                              
 \def\ints@@@{\iflimits@\negintic@
   \mathop{\intic@\multintlimits@}\limits                             
  \else\multint@\nolimits\fi                                          
  \eat@}
 \else                                                                
 \def\ints@@@{\iflimits@\negintic@
  \mathop{\intic@\multintlimits@}\limits\else
  \multint@\nolimits\fi}\fi\ints@@@}%
\def\intkern@{\mathchoice{\!\!\!}{\!\!}{\!\!}{\!\!}}%
\def\plaincdots@{\mathinner{\cdotp\cdotp\cdotp}}%
\def\intdots@{\mathchoice{\plaincdots@}%
 {{\cdotp}\mkern1.5mu{\cdotp}\mkern1.5mu{\cdotp}}%
 {{\cdotp}\mkern1mu{\cdotp}\mkern1mu{\cdotp}}%
 {{\cdotp}\mkern1mu{\cdotp}\mkern1mu{\cdotp}}}%
\def\RIfM@{\relax\protect\ifmmode}
\def\text{\RIfM@\expandafter\text@\else\expandafter\mbox\fi}
\let\nfss@text\text
\def\text@#1{\mathchoice
   {\textdef@\displaystyle\f@size{#1}}%
   {\textdef@\textstyle\tf@size{\firstchoice@false #1}}%
   {\textdef@\textstyle\sf@size{\firstchoice@false #1}}%
   {\textdef@\textstyle \ssf@size{\firstchoice@false #1}}%
   \glb@settings}

\def\textdef@#1#2#3{\hbox{{%
                    \everymath{#1}%
                    \let\f@size#2\selectfont
                    #3}}}
\newif\iffirstchoice@
\firstchoice@true
\def\Let@{\relax\iffalse{\fi\let\\=\cr\iffalse}\fi}%
\def\vspace@{\def\vspace##1{\crcr\noalign{\vskip##1\relax}}}%
\def\multilimits@{\bgroup\vspace@\Let@
 \baselineskip\fontdimen10 \scriptfont\tw@
 \advance\baselineskip\fontdimen12 \scriptfont\tw@
 \lineskip\thr@@\fontdimen8 \scriptfont\thr@@
 \lineskiplimit\lineskip
 \vbox\bgroup\ialign\bgroup\hfil$\m@th\scriptstyle{##}$\hfil\crcr}%
\def\Sb{_\multilimits@}%
\def\endSb{\crcr\egroup\egroup\egroup}%
\def\Sp{^\multilimits@}%

\newdimen\ex@
\def\rightarrowfill@#1{$#1\m@th\mathord-\mkern-6mu\cleaders
 \hbox{$#1\mkern-2mu\mathord-\mkern-2mu$}\hfill
 \mkern-6mu\mathord\rightarrow$}%
\def\leftarrowfill@#1{$#1\m@th\mathord\leftarrow\mkern-6mu\cleaders
 \hbox{$#1\mkern-2mu\mathord-\mkern-2mu$}\hfill\mkern-6mu\mathord-$}%
\def\leftrightarrowfill@#1{$#1\m@th\mathord\leftarrow
\mkern-6mu\cleaders
 \hbox{$#1\mkern-2mu\mathord-\mkern-2mu$}\hfill
 \mkern-6mu\mathord\rightarrow$}%
\def\overrightarrow{\mathpalette\overrightarrow@}%
\def\overrightarrow@#1#2{\vbox{\ialign{##\crcr\rightarrowfill@#1\crcr
 \noalign{\kern-\ex@\nointerlineskip}$\m@th\hfil#1#2\hfil$\crcr}}}%

\def\overleftarrow{\mathpalette\overleftarrow@}%
\def\overleftarrow@#1#2{\vbox{\ialign{##\crcr\leftarrowfill@#1\crcr
 \noalign{\kern-\ex@\nointerlineskip}$\m@th\hfil#1#2\hfil$\crcr}}}%
\def\overleftrightarrow{\mathpalette\overleftrightarrow@}%
\def\overleftrightarrow@#1#2{\vbox{\ialign{##\crcr
   \leftrightarrowfill@#1\crcr
 \noalign{\kern-\ex@\nointerlineskip}$\m@th\hfil#1#2\hfil$\crcr}}}%
\def\underrightarrow{\mathpalette\underrightarrow@}%
\def\underrightarrow@#1#2{\vtop{\ialign{##\crcr$\m@th\hfil#1#2\hfil
  $\crcr\noalign{\nointerlineskip}\rightarrowfill@#1\crcr}}}%

\def\underleftarrow{\mathpalette\underleftarrow@}%
\def\underleftarrow@#1#2{\vtop{\ialign{##\crcr$\m@th\hfil#1#2\hfil
  $\crcr\noalign{\nointerlineskip}\leftarrowfill@#1\crcr}}}%
\def\underleftrightarrow{\mathpalette\underleftrightarrow@}%
\def\underleftrightarrow@#1#2{\vtop{\ialign{##\crcr$\m@th
  \hfil#1#2\hfil$\crcr
 \noalign{\nointerlineskip}\leftrightarrowfill@#1\crcr}}}%

\def\qopnamewl@#1{\mathop{\operator@font#1}\nlimits@}
\let\nlimits@\displaylimits
\def\setboxz@h{\setbox\z@\hbox}

\def\varlim@#1#2{\mathop{\vtop{\ialign{##\crcr
 \hfil$#1\m@th\operator@font lim$\hfil\crcr
 \noalign{\nointerlineskip}#2#1\crcr
 \noalign{\nointerlineskip\kern-\ex@}\crcr}}}}

 \def\rightarrowfill@#1{\m@th\setboxz@h{$#1-$}\ht\z@\z@
  $#1\copy\z@\mkern-6mu\cleaders
  \hbox{$#1\mkern-2mu\box\z@\mkern-2mu$}\hfill
  \mkern-6mu\mathord\rightarrow$}
\def\leftarrowfill@#1{\m@th\setboxz@h{$#1-$}\ht\z@\z@
  $#1\mathord\leftarrow\mkern-6mu\cleaders
  \hbox{$#1\mkern-2mu\copy\z@\mkern-2mu$}\hfill
  \mkern-6mu\box\z@$}

\def\projlim{\qopnamewl@{proj\,lim}}
\def\injlim{\qopnamewl@{inj\,lim}}
\def\varinjlim{\mathpalette\varlim@\rightarrowfill@}
\def\varprojlim{\mathpalette\varlim@\leftarrowfill@}
\def\varliminf{\mathpalette\varliminf@{}}
\def\varliminf@#1{\mathop{\underline{\vrule\@depth.2\ex@\@width\z@
   \hbox{$#1\m@th\operator@font lim$}}}}
\def\varlimsup{\mathpalette\varlimsup@{}}
\def\varlimsup@#1{\mathop{\overline
  {\hbox{$#1\m@th\operator@font lim$}}}}

\def\tfrac#1#2{{\textstyle {#1 \over #2}}}%
\begingroup \catcode `|=0 \catcode `[= 1
\catcode`]=2 \catcode `\{=12 \catcode `\}=12
\catcode`\\=12 
|gdef|@alignverbatim#1\end{align}[#1|end[align]]
|gdef|@salignverbatim#1\end{align*}[#1|end[align*]]

|gdef|@alignatverbatim#1\end{alignat}[#1|end[alignat]]
|gdef|@salignatverbatim#1\end{alignat*}[#1|end[alignat*]]

|gdef|@xalignatverbatim#1\end{xalignat}[#1|end[xalignat]]
|gdef|@sxalignatverbatim#1\end{xalignat*}[#1|end[xalignat*]]

|gdef|@gatherverbatim#1\end{gather}[#1|end[gather]]
|gdef|@sgatherverbatim#1\end{gather*}[#1|end[gather*]]

|gdef|@gatherverbatim#1\end{gather}[#1|end[gather]]
|gdef|@sgatherverbatim#1\end{gather*}[#1|end[gather*]]

|gdef|@multilineverbatim#1\end{multiline}[#1|end[multiline]]
|gdef|@smultilineverbatim#1\end{multiline*}[#1|end[multiline*]]

|gdef|@arraxverbatim#1\end{arrax}[#1|end[arrax]]
|gdef|@sarraxverbatim#1\end{arrax*}[#1|end[arrax*]]

|gdef|@tabulaxverbatim#1\end{tabulax}[#1|end[tabulax]]
|gdef|@stabulaxverbatim#1\end{tabulax*}[#1|end[tabulax*]]

|endgroup

\def\align{\@verbatim \frenchspacing\@vobeyspaces \@alignverbatim
You are using the "align" environment in a style in which it is not defined.}

\@namedef{align*}{\@verbatim\@salignverbatim
You are using the "align*" environment in a style in which it is not defined.}
\expandafter\let\csname endalign*\endcsname =\endtrivlist

\def\alignat{\@verbatim \frenchspacing\@vobeyspaces \@alignatverbatim
You are using the "alignat" environment in a style in which it is not defined.}

\@namedef{alignat*}{\@verbatim\@salignatverbatim
You are using the "alignat*" environment in a style in which it is not defined.}
\expandafter\let\csname endalignat*\endcsname =\endtrivlist

\def\xalignat{\@verbatim \frenchspacing\@vobeyspaces \@xalignatverbatim
You are using the "xalignat" environment in a style in which it is not defined.}

\@namedef{xalignat*}{\@verbatim\@sxalignatverbatim
You are using the "xalignat*" environment in a style in which it is not defined.}
\expandafter\let\csname endxalignat*\endcsname =\endtrivlist

\def\gather{\@verbatim \frenchspacing\@vobeyspaces \@gatherverbatim
You are using the "gather" environment in a style in which it is not defined.}

\@namedef{gather*}{\@verbatim\@sgatherverbatim
You are using the "gather*" environment in a style in which it is not defined.}
\expandafter\let\csname endgather*\endcsname =\endtrivlist

\def\multiline{\@verbatim \frenchspacing\@vobeyspaces \@multilineverbatim
You are using the "multiline" environment in a style in which it is not defined.}

\@namedef{multiline*}{\@verbatim\@smultilineverbatim
You are using the "multiline*" environment in a style in which it is not defined.}
\expandafter\let\csname endmultiline*\endcsname =\endtrivlist

\def\arrax{\@verbatim \frenchspacing\@vobeyspaces \@arraxverbatim
You are using a type of "array" construct that is only allowed in AmS-LaTeX.}

\def\tabulax{\@verbatim \frenchspacing\@vobeyspaces \@tabulaxverbatim
You are using a type of "tabular" construct that is only allowed in AmS-LaTeX.}

\@namedef{arrax*}{\@verbatim\@sarraxverbatim
You are using a type of "array*" construct that is only allowed in AmS-LaTeX.}
\expandafter\let\csname endarrax*\endcsname =\endtrivlist

\@namedef{tabulax*}{\@verbatim\@stabulaxverbatim
You are using a type of "tabular*" construct that is only allowed in AmS-LaTeX.}
\expandafter\let\csname endtabulax*\endcsname =\endtrivlist

\def\@@eqncr{\let\@tempa\relax
    \ifcase\@eqcnt \def\@tempa{& & &}\or \def\@tempa{& &}%
      \else \def\@tempa{&}\fi
     \@tempa
     \if@eqnsw
        \iftag@
           \@taggnum
        \else
           \@eqnnum\stepcounter{equation}%
        \fi
     \fi
     \global\tag@false
     \global\@eqnswtrue
     \global\@eqcnt\z@\cr}

 \def\endequation{%
     \ifmmode\ifinner 
      \iftag@
        \addtocounter{equation}{-1} 
        $\hfil
           \displaywidth\linewidth\@taggnum\egroup \endtrivlist
        \global\tag@false
        \global\@ignoretrue   
      \else
        $\hfil
           \displaywidth\linewidth\@eqnnum\egroup \endtrivlist
        \global\tag@false
        \global\@ignoretrue 
      \fi
     \else   
      \iftag@
        \addtocounter{equation}{-1} 
        \eqno \hbox{\@taggnum}
        \global\tag@false%
        $$\global\@ignoretrue
      \else
        \eqno \hbox{\@eqnnum}
        $$\global\@ignoretrue
      \fi
     \fi\fi
 } 

 \newif\iftag@ \tag@false
 
 \def\tag{\@ifnextchar*{\@tagstar}{\@tag}}
 \def\@tag#1{%
     \global\tag@true
     \global\def\@taggnum{(#1)}}
 \def\@tagstar*#1{%
     \global\tag@true
     \global\def\@taggnum{#1}%
}

\makeatother

\begin{document}

\TeXButton{TeX field}
{
\begin{center}
\Large\textbf{Fermions in spherical field theory}
\\
\vspace{12pt}
\normalsize\textrm {Dean Lee\footnote{Supported by
the National Science Foundation under Grant 5-22968}\footnote{email: 
dlee@het.phast.umass.edu}}
\\
\small{University of Massachusetts 
\\Amherst, MA 01003}
\\
\vspace{24pt} 
\small
\parbox{360pt}{We derive the spherical field formalism for fermions. We find 
that the spherical field method is free from certain difficulties which complicate 
lattice calculations, such as fermion doubling, missing axial anomalies, 
and computational problems regarding internal fermion loops.

[PACS numbers: 11.10.Kk, 11.15Tk]}
\vspace{10pt}
\end{center}}

\section{Overview}

Spherical field theory is a new non-perturbative method for studying quantum
field theory. It was introduced in \cite{lee} and was used to describe the
interactions of scalar boson fields. In this paper we show how to extend the
spherical field method to fermionic systems.

The central idea of spherical field theory is to treat a $d$-dimensional
system as a set of coupled one-dimensional systems. This is done by
expanding field configurations of the functional integral in terms of
spherical partial waves. Regarding each partial wave as a distinct field in
a new one-dimensional theory, we interpret the functional integral as a
time-evolution equation, where the radial distance in the original theory
serves as the time parameter. For a purely bosonic system the time-evolution
equation corresponds with a multidimensional partial differential equation.
In the case of a purely fermionic system, we find that the time evolution is
described by a system of first-order ordinary differential equations. In
future work we will study mixed systems with both bosons and fermions which
are described by coupled partial differential equations.

Unlike lattice methods, spherical field theory yields an expansion which, at
any order, corresponds with a continuous system. It is therefore able to
avoid problems associated with discrete approximation methods.$\footnote{%
The author credits Robert Shrock for pointing this out.}$ There is no
doubling of fermion states, and we find the correct axial anomaly.
Furthermore internal fermion loops present no special computational
difficulties and is included in the dynamics of the time-evolution equation.
Detailed examples of such calculations will be presented in a forthcoming
paper. We anticipate that spherical field methods will be useful in the
study of non-perturbative fermionic systems, especially chiral fermions and
phenomena related to fermion loop processes.

The organization of this paper is as follows. We begin with a brief
description of Grassmanian path integrals and the fermionic analog of the
Feynman-Kac formula. We then generate the spherical expansion for free
fermion theory in two Euclidean dimensions with sources and derive the
spherical field time-evolution operator and generating functional. By
functional differentiation with respect to the sources, we obtain the
spherical field formalism for general interacting theories. Next we check
that spherical field theory produces the correct axial anomaly. We then show
how to write the time-evolution equation as a matrix system and comment on
the utility of spherical field methods in studying fermionic systems.
Although our analysis is done in two dimensions, the extension to higher
dimensions is straightforward.

\section{Grassmanian path integrals}

Let $\bar{\psi}_i(t)$ and $\psi _j(t)$ be Grassman-valued functions where $%
i,j=1,\cdots ,N$. Let $V(\bar{\psi}_i,\psi _j,t)$ be a polynomial in $\bar{%
\psi}_i$ and $\psi _j,$ ordered such that all $\psi $'s are placed on the
right and all $\bar{\psi}$'s are placed on the left. In \cite{sop} it is
shown that

\begin{eqnarray}
&&Tr\left[ T\exp \left\{ -\int_{t_I}^{t_F}dt\,V(a_i^{+},a_j^{-},t)\right\}
\right]  \label{mz} \\
&\propto &\int d\Psi d\bar{\Psi}\int\Sb \psi (t_I)=-\psi (t_F)=\Psi  \\ \bar{%
\psi}(t_I)=-\bar{\psi}(t_F)=\bar{\Psi}  \endSb \mathcal{D}\psi \,\mathcal{D}%
\bar{\psi}\exp \left\{ -\int_{t_I}^{t_F}dt\,\left[ 
\begin{array}{c}
\sum_{k=1}^N\bar{\psi}_k\tfrac{d\psi _k}{dt} \\ 
+V(\bar{\psi}_i,\psi _j,t)
\end{array}
\right] \right\} ,  \nonumber
\end{eqnarray}
where the trace is performed over the space spanned by vectors of the form 
\begin{equation}
\left| s_1\cdots s_N\right\rangle \qquad s_1=0,1;\cdots s_N=0,1;  \label{no9}
\end{equation}
and 
\begin{eqnarray}
\left\langle s_1^{\prime }\cdots s_N^{\prime }\right. \left| s_1\cdots
s_N\right\rangle &=&\delta _{s_1^{\prime }s_1}\cdots \delta _{s_N^{\prime
}s_N}  \label{mp} \\
a_i^{+}\left| s_1\cdots s_i\cdots s_N\right\rangle &=&(-1)^{s_1+\cdots
+s_{i-1}}\delta _{s_i,0}\left| s_1\cdots 1\cdots s_N\right\rangle  \nonumber
\\
a_i^{-}\left| s_1\cdots s_i\cdots s_N\right\rangle &=&(-1)^{s_1+\cdots
+s_{i-1}}\delta _{s_i,1}\left| s_1\cdots 0\cdots s_N\right\rangle . 
\nonumber
\end{eqnarray}
This is the fermionic version of the Feynman-Kac formula. A more recent
derivation using fermionic coherent states can be found in \cite{nieu} and 
\cite{mont}. The antiperiodic boundary conditions imposed at $t_I$ and $t_F$
follow as a consequence of computing the trace. We note that these are in
fact special conditions. More general boundary constraints produce
ambiguities which depend on the specific discrete approximation used to
obtain the continuum limit.

It is not clear that such antiperiodic boundary conditions can be
generalized in a coordinate-independent manner for functional integrals over
higher dimensional regions. The rigorous theory of Grassmanian functional
integration has not developed to the point where we can answer such
questions. Nevertheless functional integration is a convenient method for
deriving useful field-theoretic results, although in a somewhat heuristic
fashion. In this analysis we use the functional integral to deduce the
spherical field formalism for fermions. Although we will be careless with
regard to boundary conditions, in the end we explicitly check that the
spherical field method produces the correct generating functional for free
field theory. By functional differentiation with respect to the external
sources, we conclude that the spherical field formalism is valid for general
interacting theories.

\section{Spherical fermions}

Let us consider Euclidean field theory in two dimensions. We will use both
cartesian and polar coordinates, 
\begin{equation}
\vec{t}=(t\cos \theta ,t\sin \theta )=(x,y).  \label{cr4}
\end{equation}
In Euclidean space the gamma matrices satisfy 
\begin{equation}
\left\{ \gamma ^i,\gamma ^j\right\} =-2\delta ^{ij},  \label{b2}
\end{equation}
and we choose the representation 
\begin{equation}
\vec{\gamma}=i\vec{\sigma}.  \label{fe1}
\end{equation}
Let us start by constructing the spherical field Hamiltonian. We first
decompose the fermion fields, 
\begin{equation}
\psi =\left[ 
\begin{array}{c}
\psi ^{\uparrow }(\vec{t}) \\ 
\psi ^{\downarrow }(\vec{t})
\end{array}
\right] =\sum_{n=0,\pm 1,\cdots }\left[ 
\begin{array}{c}
\tfrac 1{\sqrt{2\pi }}\psi _n^{\uparrow }(t)e^{in\theta } \\ 
\tfrac 1{\sqrt{2\pi }}\psi _n^{\downarrow }(t)e^{in\theta }
\end{array}
\right] ,  \label{kf8}
\end{equation}
\begin{equation}
\bar{\psi}=\left[ 
\begin{array}{cc}
\bar{\psi}^{\uparrow }(\vec{t}) & \bar{\psi}^{\downarrow }(\vec{t})
\end{array}
\right] =\sum_{n=0,\pm 1,\cdots }\left[ 
\begin{array}{cc}
\frac 1{\sqrt{2\pi }}\bar{\psi}_n^{\uparrow }(t)e^{in\theta } & \frac 1{%
\sqrt{2\pi }}\bar{\psi}_n^{\downarrow }(t)e^{in\theta }
\end{array}
\right] .  \label{ov7}
\end{equation}
Using 
\begin{equation}
\vec{\sigma}\cdot \vec{\nabla}=\left[ 
\begin{array}{cc}
0 & \tfrac \partial {\partial x}-i\tfrac \partial {\partial y} \\ 
\tfrac \partial {\partial x}+i\tfrac \partial {\partial y} & 0
\end{array}
\right] =\left[ 
\begin{array}{cc}
0 & e^{-i\theta }\left( \tfrac \partial {\partial t}-\tfrac it\tfrac
\partial {\partial \theta }\right) \\ 
e^{i\theta }\left( \tfrac \partial {\partial t}+\tfrac it\tfrac \partial
{\partial \theta }\right) & 0
\end{array}
\right] ,  \label{pg5}
\end{equation}
we have 
\begin{equation}
\vec{\sigma}\cdot \vec{\nabla}\left[ 
\begin{array}{c}
\frac 1{\sqrt{2\pi }}\psi _n^{\uparrow }(t)e^{in\theta } \\ 
\frac 1{\sqrt{2\pi }}\psi _n^{\downarrow }(t)e^{in\theta }
\end{array}
\right] =\tfrac 1{\sqrt{2\pi }}\left[ 
\begin{array}{c}
\left( \tfrac{\partial \psi _n^{\downarrow }}{\partial t}+\frac nt\psi
_n^{\downarrow }\right) e^{i(n-1)\theta } \\ 
\left( \tfrac{\partial \psi _n^{\uparrow }}{\partial t}-\frac nt\psi
_n^{\uparrow }\right) e^{i(n+1)\theta }
\end{array}
\right] .  \label{fr2}
\end{equation}
The Euclidean action for free field theory with external sources, $\eta $
and $\bar{\eta}$, is 
\begin{equation}
S=-i\int d\theta dt\,t\left( \bar{\psi}(i\vec{\gamma}\cdot \vec{\nabla}%
-m)\psi +\bar{\psi}\eta +\bar{\eta}\psi \right) .  \label{xz1}
\end{equation}
In terms of partial waves,$\footnote{%
We expand $\eta $ and $\bar{\eta}$ into partial waves in the same manner as $%
\psi $ and $\bar{\psi}.$}$ 
\begin{equation}
S=-i\int dt\,t\sum_{n=0,\pm 1,\cdots }\left[ 
\begin{array}{c}
-\bar{\psi}_{-n+1}^{\uparrow }\left( \tfrac{\partial \psi _n^{\downarrow }}{%
\partial t}+\frac nt\psi _n^{\downarrow }\right) -\bar{\psi}%
_{-n-1}^{\downarrow }\left( \tfrac{\partial \psi _n^{\uparrow }}{\partial t}%
-\frac nt\psi _n^{\uparrow }\right) \\ 
-m\left( \bar{\psi}_{-n}^{\uparrow }\psi _n^{\uparrow }+\bar{\psi}%
_{-n}^{\downarrow }\psi _n^{\downarrow }\right) \\ 
+\bar{\psi}_{-n}^{\uparrow }\eta _n^{\uparrow }+\bar{\psi}_{-n}^{\downarrow
}\eta _n^{\downarrow }+\bar{\eta}_{-n}^{\uparrow }\psi _n^{\uparrow }+\bar{%
\eta}_{-n}^{\downarrow }\psi _n^{\downarrow }
\end{array}
\right] .  \label{hh}
\end{equation}
The generating functional is therefore 
\begin{equation}
\int \left( \prod_{i,n}\mathcal{D}\psi _n^i\,\mathcal{D}\bar{\psi}%
_n^i\right) \exp \left\{ \int_0^\infty dt\,t\sum_{n=0,\pm 1,\cdots
}G_n\right\} ,  \label{v3w}
\end{equation}
where $G_n$ is defined as 
\begin{eqnarray}
&&-\bar{\psi}_{-n}^{\uparrow }\left( \tfrac{\partial \psi _{n+1}^{\downarrow
}}{\partial t}+\tfrac{n+1}t\psi _{n+1}^{\downarrow }\right) -\bar{\psi}%
_{-n-1}^{\downarrow }\left( \tfrac{\partial \psi _n^{\uparrow }}{\partial t}%
-\tfrac nt\psi _n^{\uparrow }\right)  \label{nzz} \\
&&-m\left( \bar{\psi}_{-n}^{\uparrow }\psi _n^{\uparrow }+\bar{\psi}%
_{-n-1}^{\downarrow }\psi _{n+1}^{\downarrow }\right) +\bar{\psi}%
_{-n}^{\uparrow }\eta _n^{\uparrow }+\bar{\psi}_{-n-1}^{\downarrow }\eta
_{n+1}^{\downarrow }+\bar{\eta}_{-n}^{\uparrow }\psi _n^{\uparrow }+\bar{\eta%
}_{-n-1}^{\downarrow }\psi _{n+1}^{\downarrow }.  \nonumber
\end{eqnarray}
If we now define $\bar{\psi}_n^{i\prime }=t\bar{\psi}_n^i$, the generating
functional is, up to an overall constant, 
\begin{equation}
\int \left( \prod_{i,n}\mathcal{D}\psi _n^i\,\mathcal{D}\bar{\psi}%
_n^{i\prime }\right) \exp \left\{ \int_0^\infty dt\,\sum_{n=0,\pm 1,\cdots
}G_n^{\prime }\right\} ,  \label{x74}
\end{equation}
where $G_n^{\prime }$ is 
\begin{eqnarray}
&&-\bar{\psi}_{-n}^{\uparrow \prime }\left( \tfrac{\partial \psi
_{n+1}^{\downarrow }}{\partial t}+\tfrac{n+1}t\psi _{n+1}^{\downarrow
}\right) -\bar{\psi}_{-n-1}^{\downarrow \prime }\left( \tfrac{\partial \psi
_n^{\uparrow }}{\partial t}-\tfrac nt\psi _n^{\uparrow }\right)  \label{ki}
\\
&&-m\left( \bar{\psi}_{-n}^{\uparrow \prime }\psi _n^{\uparrow }+\bar{\psi}%
_{-n-1}^{\downarrow \prime }\psi _{n+1}^{\downarrow }\right) +\bar{\psi}%
_{-n}^{\uparrow \prime }\eta _n^{\uparrow }+\bar{\psi}_{-n-1}^{\downarrow
\prime }\eta _{n+1}^{\downarrow }+t\bar{\eta}_{-n}^{\uparrow }\psi
_n^{\uparrow }+t\bar{\eta}_{-n-1}^{\downarrow }\psi _{n+1}^{\downarrow }. 
\nonumber
\end{eqnarray}

Our goal is to find an equivalent expression for (\ref{x74}), in analogy
with the Feynman-Kac formula (\ref{mz})$.$ We start by defining a linear
vector space. For each finite subset 
\begin{equation}
S\subset \left\{ s_n^i\left| 
\begin{array}{c}
n=0,\pm 1,\pm 2,\cdots \\ 
i=\downarrow ,\uparrow
\end{array}
\right. \right\}  \label{ns8}
\end{equation}
we assign a vector $\left| S\right\rangle $ satisfying the following
orthogonality and normalization conditions, 
\begin{equation}
\left\langle S^{\prime }|S\right\rangle =\left\{ 
\begin{array}{cc}
0 & \text{if }S\neq S^{\prime } \\ 
1 & \text{if }S=S^{\prime }.
\end{array}
\right.  \label{u6}
\end{equation}
For later convenience we define a lexicographic order, namely, 
\begin{equation}
s_n^i<s_{n^{\prime }}^{i^{\prime }}  \label{cj7}
\end{equation}
if and only if 
\begin{eqnarray}
n &<&n^{\prime }  \label{xro} \\
\text{\ or }n &=&n^{\prime },\text{\ }i=\,\downarrow \text{,\ and\ }%
i^{\prime }=\,\uparrow \text{.}  \nonumber
\end{eqnarray}
Let $\Sigma $ be the linear space spanned by all such vectors $\left|
S\right\rangle .$ Let us define operators $a_n^{i+}$ and $a_n^{i-}$ by the
following relations, 
\begin{eqnarray}
a_n^{i-}\left| S\right\rangle &=&\left\{ 
\begin{array}{cc}
0 & \text{if }s_n^i\notin S \\ 
(-1)^{\#}\left| S-s_n^i\right\rangle & \text{if }s_n^i\in S,
\end{array}
\right.  \label{xye} \\
a_n^{i+}\left| S\right\rangle &=&\left\{ 
\begin{array}{cc}
0 & \text{if }s_n^i\in S \\ 
(-1)^{\#}\left| S\cup s_n^i\right\rangle & \text{if }s_n^i\notin S,
\end{array}
\right.  \nonumber
\end{eqnarray}
where \# is the number of elements in $S$ which are less than $s_n^i$.
Comparing (\ref{x74}) with (\ref{mz}), we make the correspondences$\footnote{%
We denote the conjugate of $a_n^{i-}$ as $a_n^{i+}.$ Although $a_n^{i-}$
corresponds with a partial wave with orbital angular momentum $n,$ we note
that $a_n^{i+}$ corresponds with a partial wave with orbital angular
momentum $-n-1$ or $-n+1$, depending on $i.$}$ 
\begin{eqnarray}
\psi _n^{\uparrow },\bar{\psi}_{-n-1}^{\downarrow \prime } &\leftrightarrow
&a_n^{\uparrow -},a_n^{\uparrow +}  \label{nx5} \\
\psi _{n+1}^{\downarrow },\bar{\psi}_{-n}^{\uparrow \prime }
&\leftrightarrow &a_{n+1}^{\downarrow -},a_{n+1}^{\downarrow +}  \nonumber
\end{eqnarray}
and define 
\begin{equation}
Z[\bar{\eta},\eta ]=\lim_{\varepsilon \rightarrow 0}\tfrac{Tr\left[ T\exp
\left\{ -\int_0^\infty dt\,\sum\limits_{n=0,\pm 1,\cdots }H_n^\varepsilon (%
\bar{\eta},\eta ,t)\right\} \right] }{Tr\left[ T\exp \left\{ -\int_0^\infty
dt\,\sum\limits_{n=0,\pm 1,\cdots }H_n^\varepsilon (0,0,t)\right\} \right] },
\label{xh}
\end{equation}
where 
\begin{eqnarray}
H_n^\varepsilon &=&\tfrac{n+1}ta_{n+1}^{\downarrow +}a_{n+1}^{\downarrow
-}-\tfrac nta_n^{\uparrow +}a_n^{\uparrow -}+(m+\varepsilon
)a_{n+1}^{\downarrow +}a_n^{\uparrow -}+(m+\varepsilon )a_n^{\uparrow
+}a_{n+1}^{\downarrow -}  \label{ebz} \\
&&-a_{n+1}^{\downarrow +}\eta _n^{\uparrow }-a_n^{\uparrow +}\eta
_{n+1}^{\downarrow }-t\bar{\eta}_{-n}^{\uparrow }a_n^{\uparrow -}-t\bar{\eta}%
_{-n-1}^{\downarrow }a_{n+1}^{\downarrow -}.  \nonumber
\end{eqnarray}
The traces in (\ref{xh}) are performed over the space $\Sigma .$ In practise
it is not necessary to explicitly compute these traces, since only the
ground state projection contributes. For $m=0$, however, the ground state is
degenerate and the $\varepsilon \rightarrow 0$ prescription picks out the
correct ground state. We will refer to (\ref{xh}) as the fermionic spherical
field ansatz. For notational ease we will suppress the $\varepsilon $ terms.
We now show that $Z[\bar{\eta},\eta ]$ is the correct generating functional
for free field theory.

We note that $\Sigma $ can be decomposed as a tensor product space with the
identification 
\begin{equation}
\left| S\right\rangle \leftrightarrow \bigotimes_{n=0,\pm 1,\cdots }\left|
S\cap \{s_n^{\uparrow },s_{n+1}^{\downarrow }\}\right\rangle .  \label{zxy}
\end{equation}
Since $H_n(\bar{\eta},\eta ,t)$ acts upon only the $n$th component in the
tensor product, we can write $Z[\bar{\eta},\eta ]$ as a product 
\begin{equation}
Z[\bar{\eta},\eta ]=\prod_{n=0,\pm 1,\cdots }Z_n[\bar{\eta}_{-n}^{\uparrow },%
\bar{\eta}_{-n-1}^{\downarrow },\eta _n^{\uparrow },\eta _{n+1}^{\downarrow
}],  \label{uj}
\end{equation}
where 
\begin{equation}
Z_n[\bar{\eta}_{-n}^{\uparrow },\bar{\eta}_{-n-1}^{\downarrow },\eta
_n^{\uparrow },\eta _{n+1}^{\downarrow }]=\tfrac{Tr\left[ T\exp \left\{
-\int_0^\infty dt\,H_n(\bar{\eta},\eta ,t)\right\} \right] }{Tr\left[ T\exp
\left\{ -\int_0^\infty dt\,H_n(0,0,t)\right\} \right] }.  \label{mr3}
\end{equation}
In (\ref{mr3}) the trace is performed over the four-dimensional space
spanned by the vectors 
\begin{equation}
\left| \emptyset \right\rangle ,\;\left| \{s_{n+1}^{\downarrow
}\}\right\rangle ,\;\left| \{s_n^{\uparrow }\}\right\rangle ,\;\left|
\{s_n^{\uparrow },s_{n+1}^{\downarrow }\}\right\rangle .  \label{ns5}
\end{equation}
In Appendix 1 we derive the result, 
\begin{equation}
Z=\exp \left\{ \sum_{n=0,\pm 1,\cdots }\int dt_1dt_2\left[ 
\begin{array}{cc}
t_1\bar{\eta}_{-n}^{\uparrow }(t_1) & t_1\bar{\eta}_{-n-1}^{\downarrow }(t_1)
\end{array}
\right] \mathbf{M}_n(t_1,t_2)\left[ 
\begin{array}{c}
t_2\eta _n^{\uparrow }(t_2) \\ 
t_2\eta _{n+1}^{\downarrow }(t_2)
\end{array}
\right] \right\} ,  \label{g4a}
\end{equation}
where the matrix $\mathbf{M}_n(t_1,t_2)$ is defined as\footnote{$I_i$ and $%
K_i$ are the $i^{th}$ order modified Bessel functions of the first and
second kinds respectively.} 
\begin{eqnarray}
&&\theta (t_1-t_2)\left[ 
\begin{array}{cc}
mK_n(\left| m\right| t_1)I_n(\left| m\right| t_2) & \left| m\right|
K_n(\left| m\right| t_1)I_{n+1}(\left| m\right| t_2) \\ 
\left| m\right| K_{n+1}(\left| m\right| t_1)I_n(\left| m\right| t_2) & 
mK_{n+1}(\left| m\right| t_1)I_{n+1}(\left| m\right| t_2)
\end{array}
\right]  \label{xu} \\
&&+\theta (t_2-t_1)\left[ 
\begin{array}{cc}
mI_n(\left| m\right| t_1)K_n(\left| m\right| t_2) & -\left| m\right|
I_n(\left| m\right| t_1)K_{n+1}(\left| m\right| t_2) \\ 
-\left| m\right| I_{n+1}(\left| m\right| t_1)K_n(\left| m\right| t_2) & 
mI_{n+1}(\left| m\right| t_1)K_{n+1}(\left| m\right| t_2)
\end{array}
\right]  \nonumber
\end{eqnarray}
for $m\neq 0,$ and

\begin{equation}
\theta (t_1-t_2)\left[ 
\begin{array}{cc}
0 & \frac{\theta (-n-\frac 12)t_1^n}{t_2^{n+1}} \\ 
\frac{\theta (n+\frac 12)t_2^n}{t_1^{n+1}} & 0
\end{array}
\right] -\theta (t_2-t_1)\left[ 
\begin{array}{cc}
0 & \frac{\theta (n+\frac 12)t_1^n}{t_2^{n+1}} \\ 
\frac{\theta (-n-\frac 12)t_2^n}{t_1^{n+1}} & 0
\end{array}
\right]  \label{rx3}
\end{equation}
for $m=0.$

Let us now compare these results with the known results for free field
theory. The two-point free field correlator is 
\begin{eqnarray}
\Delta ^{ij}(\vec{t}) &=&\int \tfrac{d^2\vec{k}}{(2\pi )^2}\tfrac{-\vec{k}%
\cdot \vec{\gamma}+m}{\vec{k}^2+m^2}e^{i\vec{k}\cdot \vec{t}}=\int \tfrac{d^2%
\vec{k}}{(2\pi )^2}\tfrac{-i\vec{k}\cdot \vec{\sigma}+m}{\vec{k}^2+m^2}e^{i%
\vec{k}\cdot \vec{t}}  \label{ci6} \\
&=&\int \tfrac{d^2\vec{k}}{(2\pi )^2}\tfrac{-\vec{\sigma}\cdot \vec{\nabla}+m%
}{\vec{k}^2+m^2}e^{i\vec{k}\cdot \vec{t}}.  \nonumber
\end{eqnarray}
Integrating over $\vec{k}$, we find 
\begin{eqnarray}
\Delta ^{ij}(\vec{t}) &=&\tfrac 1{2\pi }\left[ mK_0(\left| m\right|
t)+\left| m\right| K_1(\left| m\right| t)\cdot (\sigma ^x\cos \theta +\sigma
^y\sin \theta )\right]  \label{e4x} \\
&=&\tfrac 1{2\pi }\left[ 
\begin{array}{cc}
mK_0(\left| m\right| t) & \left| m\right| e^{-i\theta }K_1(\left| m\right| t)
\\ 
\left| m\right| e^{i\theta }K_1(\left| m\right| t) & mK_0(\left| m\right| t)
\end{array}
\right]  \nonumber
\end{eqnarray}
for $m\neq 0.$ When $m\rightarrow 0$ we find 
\begin{equation}
\Delta ^{ij}(\vec{t})=\tfrac 1{2\pi }\left[ 
\begin{array}{cc}
0 & e^{-i\theta }\frac 1t \\ 
e^{i\theta }\frac 1t & 0
\end{array}
\right] .  \label{3z}
\end{equation}
From (\ref{e4x}) and (\ref{3z}) it is straightforward to recover the
spherical correlation functions in (\ref{xu}) and (\ref{rx3}). We conclude
that the fermionic spherical field ansatz produces the correct generating
functional for free fermions. These correlation functions are part of the
spherical Feynman rules for fermions. For future reference we have written
these in a more convenient format in Appendix 2. By functional
differentiation with respect to the sources $\eta $ and $\bar{\eta}$, we
conclude that the spherical field formalism is valid for general interacting
theories.

\section{Axial anomaly}

We now show that the spherical field formalism yields the correct form for
the axial anomaly. We consider free massless fermions, again in two
Euclidean dimensions. Let us define 
\begin{equation}
S^{\mu \nu }(\vec{t})=\left\langle 0\left| V^\mu (\vec{t})A^\nu (0)\right|
0\right\rangle _E,  \label{aa6}
\end{equation}
where $V^\mu $ and $A^\mu $ are the vector and axial vector currents, and
the subscript $E$ is intended as a reminder that our Euclidean correlation
function is defined as the analytic continuation of the corresponding
time-ordered function in Minkowski space.\footnote{%
Euclidean correlation functions can also be defined without analytic
continuation. However this involves new complications which are described in 
\cite{ost}.} We note that 
\begin{equation}
\left\langle 0\left| V^\mu (\vec{t})\partial _\nu A^\nu (0)\right|
0\right\rangle _E=-\partial _\nu ^{\vec{t}}\left\langle 0\left| V^\mu (\vec{t%
})A^\nu (0)\right| 0\right\rangle _E=-\partial _\nu S^{\mu \nu }.  \label{a7}
\end{equation}
The one-loop process corresponding with $S^{\mu \nu }$ carries a logarithmic
divergence proportional to $\varepsilon ^{\mu \nu },$ and the regulated
value of $S^{\mu \nu }$ will depend on our definitions of $V^\mu $ and $%
A^\mu $ as operator products$.$ In our discussion here we will remove this
ambiguity by considering the symmetric combination $S^{\mu \nu }+S^{\nu \mu }
$. Let us define vector and axial vector currents,

\begin{equation}
V^\mu (\vec{t})=\bar{\psi}(\vec{t})\gamma ^\mu \psi (\vec{t})=i\bar{\psi}(%
\vec{t})\sigma ^\mu \psi (\vec{t})  \label{a1}
\end{equation}
\begin{equation}
A^\mu (\vec{t})=\bar{\psi}(\vec{t})\gamma ^\mu \gamma _5\psi (\vec{t})=i\bar{%
\psi}(\vec{t})\sigma ^\mu \sigma ^z\psi (\vec{t}).  \label{a2}
\end{equation}
Expanding the currents in terms of partial waves, we have 
\begin{equation}
V^\mu (\vec{t})=\tfrac 1{2\pi }\sum_{k,n=0,\pm 1,\cdots }e^{i(k-n)\theta
}\left[ v_{\uparrow \downarrow }^\mu \bar{\psi}_{-n}^{\uparrow }(t)\psi
_k^{\downarrow }(t)+v_{\downarrow \uparrow }^\mu \bar{\psi}_{-n}^{\downarrow
}(t)\psi _k^{\uparrow }(t)\right]   \label{a3}
\end{equation}
\begin{equation}
A^\mu (\vec{t})=\tfrac 1{2\pi }\sum_{k,n=0,\pm 1,\cdots }e^{i(k-n)\theta
}\left[ a_{\uparrow \downarrow }^\mu \bar{\psi}_{-n}^{\uparrow }(t)\psi
_k^{\downarrow }(t)+a_{\downarrow \uparrow }^\mu \bar{\psi}_{-n}^{\downarrow
}(t)\psi _k^{\uparrow }(t)\right] ,  \label{a4}
\end{equation}
where 
\begin{equation}
v_{\uparrow \downarrow }^1=i,\;v_{\downarrow \uparrow }^1=i,\;v_{\uparrow
\downarrow }^2=1,\;v_{\downarrow \uparrow }^2=-1  \label{a5}
\end{equation}
\begin{equation}
a_{\uparrow \downarrow }^1=-i,\;a_{\downarrow \uparrow }^1=i,\;a_{\uparrow
\downarrow }^2=-1,\;a_{\downarrow \uparrow }^2=-1.  \label{a6}
\end{equation}
From (\ref{a3}) and (\ref{a4}), we have 
\begin{equation}
S^{\mu \nu }(\vec{t})=\tfrac 1{\left( 2\pi \right) ^2}\left\langle 0\left|
\left[ 
\begin{array}{c}
e^{2i\theta }v_{\uparrow \downarrow }^\mu \bar{\psi}_1^{\uparrow }(t)\psi
_1^{\downarrow }(t) \\ 
+e^{-2i\theta }v_{\downarrow \uparrow }^\mu \bar{\psi}_{-1}^{\downarrow
}(t)\psi _{-1}^{\uparrow }(t)
\end{array}
\right] \left[ 
\begin{array}{c}
a_{\uparrow \downarrow }^\nu \bar{\psi}_0^{\uparrow }(0)\psi _0^{\downarrow
}(0) \\ 
+a_{\downarrow \uparrow }^\nu \bar{\psi}_0^{\downarrow }(0)\psi _0^{\uparrow
}(0)
\end{array}
\right] \right| 0\right\rangle _E.  \label{aa7}
\end{equation}
Recalling the correlator results from the previous section, we have 
\begin{equation}
S^{\mu \nu }(\vec{t})=\tfrac 1{\left( 2\pi \right) ^2t^2}\left( e^{2i\theta
}v_{\uparrow \downarrow }^\mu a_{\uparrow \downarrow }^\nu +e^{-2i\theta
}v_{\downarrow \uparrow }^\mu a_{\downarrow \uparrow }^\nu \right) ,
\label{a8}
\end{equation}
and so 
\begin{eqnarray}
-\partial _\nu S^{\mu \nu }-\partial _\nu S^{\nu \mu } &=&-\tfrac{4i}{\left(
2\pi \right) ^2}\delta ^{\mu 1}\left[ \tfrac \partial {\partial x}\left( 
\tfrac{2xy}{\left( x^2+y^2\right) ^2}\right) -\tfrac \partial {\partial
y}\left( \tfrac{x^2-y^2}{\left( x^2+y^2\right) ^2}\right) \right] 
\label{aa9} \\
&&+\tfrac{4i}{\left( 2\pi \right) ^2}\delta ^{\mu 2}\left[ \tfrac \partial
{\partial x}\left( \tfrac{x^2-y^2}{\left( x^2+y^2\right) ^2}\right) +\tfrac
\partial {\partial y}\left( \tfrac{2xy}{\left( x^2+y^2\right) ^2}\right)
\right] .  \nonumber
\end{eqnarray}
Integrating with any smooth test function we recognize these terms as
derivatives of the Dirac delta distribution, 
\begin{eqnarray}
-\partial _\nu S^{\mu \nu }-\partial _\nu S^{\nu \mu } &=&-\tfrac{4i}{\left(
2\pi \right) ^2}\delta ^{\mu 1}\left[ -\pi \partial _2\delta (\vec{t}%
)\right] +\tfrac{4i}{\left( 2\pi \right) ^2}\delta ^{\mu 2}\left[ -\pi
\partial _1\delta (\vec{t})\right]   \label{a10} \\
&=&\tfrac i\pi \varepsilon ^{\mu \nu }\partial _\nu \delta (\vec{t}). 
\nonumber
\end{eqnarray}
We conclude that
\begin{equation}
\int d^2t\,e^{i\vec{p}\cdot \vec{t}}\left[ \left\langle 0\left| V^\mu (\vec{t%
})\partial _\nu A^\nu (0)\right| 0\right\rangle _E-\left\langle 0\left|
\partial _\nu V^\nu (\vec{t})A^\mu (0)\right| 0\right\rangle _E\right]
=\tfrac 1\pi \varepsilon ^{\mu \nu }p_\nu .  \label{zhr}
\end{equation}
If we now choose to maintain conservation of the vector current, we have
\begin{equation}
\int d^2t\,e^{i\vec{p}\cdot \vec{t}}\left\langle 0\left| V^\mu (\vec{t}%
)\partial _\nu A^\nu (0)\right| 0\right\rangle _E=\tfrac 1\pi \varepsilon
^{\mu \nu }p_\nu ,  \label{a12}
\end{equation}
which is the desired result for the axial anomaly (see \cite{nin}). This
should not be surprising. With $\partial _\nu A^\nu $ placed at the origin,
the calculation we have done is the same as that of standard field theory in
position space. For the case when $\partial _\nu A^\nu $ is not at the
origin, our spherical field calculation generates an expansion of 
\begin{equation}
S^{\mu \nu }(\vec{t}-\vec{t}^{\prime })=\left\langle 0\left| V^\mu (\vec{t}%
)A^\nu (\vec{t}^{\prime })\right| 0\right\rangle _E  \label{ydf}
\end{equation}
in terms of sums of spherical waves. The partial sums of this expansion
converge pointwise (except at $\vec{t}=\vec{t}^{\prime })$ to the result (%
\ref{a8}), and we again get the correct axial anomaly.

\section{Matrix representation}

Spherical fermion fields can be studied using ordinary matrices. Unlike the
spherical bosonic system which corresponds with a multidimensional partial
differential equation, the spherical fermionic system corresponds with a set
of coupled first-order ordinary differential equations. We illustrate some
basic methods here using the free fermion system.

Let us define the following column vectors 
\begin{equation}
\left[ 
\begin{array}{c}
0 \\ 
0 \\ 
0 \\ 
1
\end{array}
\right] =\left| \emptyset \right\rangle ,\quad \left[ 
\begin{array}{c}
0 \\ 
0 \\ 
1 \\ 
0
\end{array}
\right] =\left| \{s_{n+1}^{\downarrow }\}\right\rangle ,\quad \left[ 
\begin{array}{c}
0 \\ 
1 \\ 
0 \\ 
0
\end{array}
\right] =\left| \{s_n^{\uparrow }\}\right\rangle ,\quad \left[ 
\begin{array}{c}
1 \\ 
0 \\ 
0 \\ 
0
\end{array}
\right] =\left| \{s_n^{\uparrow },s_{n+1}^{\downarrow }\}\right\rangle .
\label{u7f}
\end{equation}
We now write the Hamiltonian and creation and annihilation operators as
matrices, 
\begin{equation}
H_n(0,0,t)=\left[ 
\begin{array}{cccc}
\frac 1t & 0 & 0 & 0 \\ 
0 & \frac{-n}t & m & 0 \\ 
0 & m & \frac{n+1}t & 0 \\ 
0 & 0 & 0 & 0
\end{array}
\right] \   \label{r34}
\end{equation}
and 
\begin{eqnarray}
a_{n+1}^{\downarrow +} &=&\left[ 
\begin{array}{cccc}
0 & -1 & 0 & 0 \\ 
0 & 0 & 0 & 0 \\ 
0 & 0 & 0 & 1 \\ 
0 & 0 & 0 & 0
\end{array}
\right] \quad a_{n+1}^{\downarrow -}=\left[ 
\begin{array}{cccc}
0 & 0 & 0 & 0 \\ 
-1 & 0 & 0 & 0 \\ 
0 & 0 & 0 & 0 \\ 
0 & 0 & 1 & 0
\end{array}
\right]  \label{rtx} \\
a_n^{\uparrow +} &=&\left[ 
\begin{array}{cccc}
0 & 0 & 1 & 0 \\ 
0 & 0 & 0 & 1 \\ 
0 & 0 & 0 & 0 \\ 
0 & 0 & 0 & 0
\end{array}
\right] \quad a_n^{\uparrow -}=\left[ 
\begin{array}{cccc}
0 & 0 & 0 & 0 \\ 
0 & 0 & 0 & 0 \\ 
1 & 0 & 0 & 0 \\ 
0 & 1 & 0 & 0
\end{array}
\right] .  \nonumber
\end{eqnarray}

As an illustrative example we calculate the correlation function 
\begin{equation}
\left\langle 0\right| \psi _1^{\downarrow }(t)\bar{\psi}_0^{\uparrow
}(0)\left| 0\right\rangle _E  \label{eg3}
\end{equation}
for the case $m\neq 0$. Let 
\begin{equation}
\mathbf{U}(t_2,t_1)=T\exp \left\{ -\int_{t_1}^{t_2}dt\,H_0(0,0,t)\right\} .
\label{nr2}
\end{equation}
Using the fact that 
\begin{equation}
\frac \partial {\partial t_2}\mathbf{U}(t_2,t_1)=-H_0(0,0,t_2)\cdot \mathbf{U%
}(t_2,t_1),  \label{mcl}
\end{equation}
we find 
\begin{equation}
\mathbf{U}(t_2,t_1)\left| \emptyset \right\rangle =\left| \emptyset
\right\rangle ,  \label{nn4}
\end{equation}
\begin{eqnarray}
\mathbf{U}(t_2,t_1)\left| \{s_1^{\downarrow }\}\right\rangle &=&\left|
m\right| t_1\cdot \left[ 
\begin{array}{c}
K_0(\left| m\right| t_1)I_1(\left| m\right| t_2) \\ 
+I_0(\left| m\right| t_1)K_1(\left| m\right| t_2)
\end{array}
\right] \left| \{s_1^{\downarrow }\}\right\rangle  \label{cc} \\
&&+mt_1\cdot \left[ 
\begin{array}{c}
-K_0(\left| m\right| t_1)I_0(\left| m\right| t_2) \\ 
+I_0(\left| m\right| t_1)K_0(\left| m\right| t_2)
\end{array}
\right] \left| \{s_0^{\uparrow }\}\right\rangle ,  \nonumber
\end{eqnarray}
\begin{eqnarray}
\mathbf{U}(t_2,t_1)\left| \{s_0^{\uparrow }\}\right\rangle &=&\left|
m\right| t_1\cdot \left[ 
\begin{array}{c}
K_1(\left| m\right| t_1)I_0(\left| m\right| t_2) \\ 
+I_1(\left| m\right| t_1)K_0(\left| m\right| t_2)
\end{array}
\right] \left| \{s_0^{\uparrow }\}\right\rangle  \label{uur} \\
&&+mt_1\cdot \left[ 
\begin{array}{c}
-K_1(\left| m\right| t_1)I_1(\left| m\right| t_2) \\ 
+I_1(\left| m\right| t_1)K_1(\left| m\right| t_2)
\end{array}
\right] \left| \{s_1^{\downarrow }\}\right\rangle ,  \nonumber
\end{eqnarray}
\begin{equation}
\mathbf{U}(t_2,t_1)\left| \{s_n^{\uparrow },s_{n+1}^{\downarrow
}\}\right\rangle =\tfrac{t_1}{t_2}\left| \{s_n^{\uparrow
},s_{n+1}^{\downarrow }\}\right\rangle .  \label{3t}
\end{equation}
Therefore 
\begin{eqnarray}
\lim\Sb t_i\rightarrow 0^{+}  \\ t_f\rightarrow \infty  \endSb Tr\,\mathbf{U}%
(t_f,t_i) &=&1+\lim\Sb t_i\rightarrow 0^{+}  \\ t_f\rightarrow \infty 
\endSb \left| m\right| t_iK_1(\left| m\right| t_i)I_0(\left| m\right| t_f)
\label{vv3} \\
&=&\lim_{t_f\rightarrow \infty }I_0(\left| m\right| t_f)  \nonumber \\
&&  \nonumber
\end{eqnarray}
and 
\begin{equation}
\lim\Sb t_i\rightarrow 0^{+}  \\ t_f\rightarrow \infty  \endSb Tr\,\mathbf{U}%
(t_f,t)a_1^{\downarrow -}\mathbf{U}(t,t_i)\tfrac 1{t_i}a_1^{\downarrow
+}=\lim_{t_f\rightarrow \infty }\left| m\right| \cdot \left[ 
\begin{array}{c}
K_1(\left| m\right| t)I_0(\left| m\right| t_f) \\ 
+I_1(\left| m\right| t)K_0(\left| m\right| t_f)
\end{array}
\right] .  \label{vxz}
\end{equation}
We conclude that 
\begin{eqnarray}
\left\langle 0\right| \psi _1^{\downarrow }(t)\bar{\psi}_0^{\uparrow
}(0)\left| 0\right\rangle _E &=&\lim\Sb t_i\rightarrow 0^{+}  \\ %
t_f\rightarrow \infty  \endSb \tfrac{Tr\,\mathbf{U}(t_f,t)a_1^{\downarrow -}%
\mathbf{U}(t,t_i)\tfrac 1{t_i}a_1^{\downarrow +}}{Tr\,\mathbf{U}(t_f,t_i)}
\label{vy9} \\
&=&\left| m\right| K_1(\left| m\right| t),  \nonumber
\end{eqnarray}
which agrees with (\ref{xu}).

Calculations in interacting theories are done in a similar manner. There we
encounter new terms in the Hamiltonian which can be found by functional
differentiation with respect to the sources $\eta $ and $\bar{\eta}.$
Several detailed examples of interacting systems will be presented in later
work.

\section{Summary}

In this paper we have extended the spherical field formalism to include
fermionic systems. Since spherical field theory deals with continuous
systems, it avoids problems associated with discrete approximation methods,
such as fermion doubling and cancelled axial anomalies. The spherical field
method should therefore be useful in studying chiral fermion systems.

We have shown that the time evolution of spherical fermion fields can be
modelled using matrices and is described by a set of coupled first-order
ordinary differential equations. We recall that in lattice field theory
Grassmanian variables are associated with each fermionic degree of freedom
at each lattice site. In spherical field theory, however, Grassmanian
creation and annihilation operators are associated with each spherical
partial wave. Although we have not analyzed interacting systems in this
work, we anticipate that for comparable levels of accuracy spherical field
theory will require manipulating much smaller anticommuting algebras. Since
large sets of anticommuting variables present serious computational
problems, spherical field theory may yield significant improvements in the
numerical calculation of fermionic interactions.

\TeXButton{vspace}{\vspace{12pt}}

\noindent \TeXButton{TeX field}
{\vspace{12pt}
\Large\bf{Acknowledgements}
\normalsize\rm}

The author is grateful to Howard Georgi for useful discussions regarding
this work and Robert Shrock for providing the impetus and encouragement to
investigate fermions using spherical field theory. The author also thanks
the Harvard physics department, where this work was started.

\TeXButton{vspace}{\vspace{12pt}}

\noindent \TeXButton{TeX field}
{\vspace{12pt}
\Large\bf{Appendix 1}
\normalsize\rm}

In this appendix we derive the results (\ref{g4a}), (\ref{xu}), and (\ref
{rx3}). Let us define 
\begin{equation}
W_n=\ln Z_n.  \label{ndt}
\end{equation}
Using the notation, 
\begin{equation}
\left\langle A\right\rangle _t\equiv \tfrac{Tr\left[ T\exp \left\{
-\int_t^\infty dt^{\prime }\,H_n(\bar{\eta},\eta ,t^{\prime })\right\}
\,A\,T\exp \left\{ -\int_0^tdt^{\prime }H_n(\bar{\eta},\eta ,t^{\prime
})\right\} \right] }{Tr\left[ T\exp \left\{ -\int_0^\infty dt^{\prime }\,H_n(%
\bar{\eta},\eta ,t^{\prime })\right\} \right] },  \label{6dx}
\end{equation}
we note that 
\begin{equation}
W_n\tfrac{\overleftarrow{\delta }}{\delta \eta _{n+1}^{\downarrow }(t)}=\ln
Z_n\tfrac{\overleftarrow{\delta }}{\delta \eta _{n+1}^{\downarrow }(t)}%
=\left\langle a_n^{\uparrow +}\right\rangle _t.  \label{b4}
\end{equation}
Differentiating with respect to $t$ we find 
\begin{equation}
\tfrac d{dt}\left[ W_n\tfrac{\overleftarrow{\delta }}{\delta \eta
_{n+1}^{\downarrow }(t)}\right] =\left\langle [H_n(\bar{\eta},\eta
,t),a_n^{\uparrow +}]\right\rangle _t=\left\langle -\tfrac nta_n^{\uparrow
+}+ma_{n+1}^{\downarrow +}-t\bar{\eta}_{-n}^{\uparrow }\right\rangle _t.
\label{gez}
\end{equation}
Differentiating again we get 
\begin{eqnarray}
&&\tfrac{d^2}{dt^2}\left[ W_n\tfrac{\overleftarrow{\delta }}{\delta \eta
_{n+1}^{\downarrow }(t)}\right]  \label{87v} \\
&=&\left\langle \tfrac n{t^2}a_n^{\uparrow +}-\bar{\eta}_{-n}^{\uparrow
}-t\tfrac d{dt}\bar{\eta}_{-n}^{\uparrow }+[H_n(\bar{\eta},\eta ,t),-\tfrac
nta_n^{\uparrow +}+ma_{n+1}^{\downarrow +}-t\bar{\eta}_{-n}^{\uparrow
}]\right\rangle _t  \nonumber \\
&=&\left\langle \left( \tfrac n{t^2}+\tfrac{n^2}{t^2}+m^2\right)
a_n^{\uparrow +}+\tfrac mta_{n+1}^{\downarrow +}+(n-1)\bar{\eta}%
_{-n}^{\uparrow }-t\tfrac d{dt}\bar{\eta}_{-n}^{\uparrow }-mt\bar{\eta}%
_{-n-1}^{\downarrow }\right\rangle _t.  \nonumber
\end{eqnarray}
We now combine (\ref{b4}), (\ref{gez}), and (\ref{87v}), 
\begin{eqnarray}
&&\left[ t^2\tfrac{d^2}{dt^2}-t\tfrac d{dt}-(n^2+2n+m^2t^2)\right] W_n\tfrac{%
\overleftarrow{\delta }}{\delta \eta _{n+1}^{\downarrow }(t)}  \label{z7z} \\
&=&\left\langle nt^2\bar{\eta}_{-n}^{\uparrow }-t^3\tfrac d{dt}\bar{\eta}%
_{-n}^{\uparrow }-mt^3\bar{\eta}_{-n-1}^{\downarrow }\right\rangle _t=nt^2%
\bar{\eta}_{-n}^{\uparrow }-t^3\tfrac d{dt}\bar{\eta}_{-n}^{\uparrow }-mt^3%
\bar{\eta}_{-n-1}^{\downarrow }.  \nonumber
\end{eqnarray}
By similar steps we find 
\begin{eqnarray}
\left[ t^2\tfrac{d^2}{dt^2}-t\tfrac d{dt}-(n^2-1+m^2t^2)\right] W_n\tfrac{%
\overleftarrow{\delta }}{\delta \eta _n^{\uparrow }(t)} &=&-(n+1)t^2\bar{\eta%
}_{-n-1}^{\downarrow }  \label{1} \\
&&-t^3\tfrac d{dt}\bar{\eta}_{-n-1}^{\downarrow }-mt^3\bar{\eta}%
_{-n}^{\uparrow },  \nonumber
\end{eqnarray}
\begin{eqnarray}
\left[ t^2\tfrac{d^2}{dt^2}+t\tfrac d{dt}-(n^2+m^2t^2)\right] \tfrac \delta
{t\cdot \delta \bar{\eta}_{-n}^{\uparrow }(t)}W_n &=&(n+1)t\eta
_{n+1}^{\downarrow }  \label{v32} \\
&&+t^2\tfrac d{dt}\eta _{n+1}^{\downarrow }-mt^2\eta _n^{\uparrow }, 
\nonumber
\end{eqnarray}
\begin{eqnarray}
\left[ t^2\tfrac{d^2}{dt^2}+t\tfrac d{dt}-((n+1)^2+m^2t^2)\right] \tfrac
\delta {t\cdot \delta \bar{\eta}_{-n-1}^{\downarrow }(t)}W_n &=&-nt\eta
_n^{\uparrow }  \label{v3z2} \\
&&+t^2\tfrac d{dt}\eta _n^{\uparrow }-mt^2\eta _{n+1}^{\downarrow }. 
\nonumber
\end{eqnarray}
For $m\neq 0$ the general solution for $W_n$ is 
\begin{eqnarray}
&&\int dt_1dt_2\left[ 
\begin{array}{cc}
t_1\bar{\eta}_{-n}^{\uparrow }(t_1) & t_1\bar{\eta}_{-n-1}^{\downarrow }(t_1)
\end{array}
\right] \mathbf{M}_n(t_1,t_2)\left[ 
\begin{array}{c}
t_2\eta _n^{\uparrow }(t_2) \\ 
t_2\eta _{n+1}^{\downarrow }(t_2)
\end{array}
\right]  \label{xu5} \\
&&+F\left[ \int dt\,tK_n(\left| m\right| t)\bar{\eta}_{-n}^{\uparrow
}(t),\cdots ,\int dt\,tI_{n+1}(\left| m\right| t)\eta _{n+1}^{\downarrow
}(t)\right] ,  \nonumber
\end{eqnarray}
where $F$ is the general homogeneous solution and $\mathbf{M}_n(t_1,t_2)$ is
given by 
\begin{eqnarray}
&&\theta (t_1-t_2)\left[ 
\begin{array}{cc}
mK_n(\left| m\right| t_1)I_n(\left| m\right| t_2) & \left| m\right|
K_n(\left| m\right| t_1)I_{n+1}(\left| m\right| t_2) \\ 
\left| m\right| K_{n+1}(\left| m\right| t_1)I_n(\left| m\right| t_2) & 
mK_{n+1}(\left| m\right| t_1)I_{n+1}(\left| m\right| t_2)
\end{array}
\right]  \label{3f} \\
&&+\theta (t_2-t_1)\left[ 
\begin{array}{cc}
mI_n(\left| m\right| t_1)K_n(\left| m\right| t_2) & -\left| m\right|
I_n(\left| m\right| t_1)K_{n+1}(\left| m\right| t_2) \\ 
-\left| m\right| I_{n+1}(\left| m\right| t_1)K_n(\left| m\right| t_2) & 
mI_{n+1}(\left| m\right| t_1)K_{n+1}(\left| m\right| t_2)
\end{array}
\right]  \nonumber
\end{eqnarray}
for $m\neq 0$, and 
\begin{equation}
\theta (t_1-t_2)\left[ 
\begin{array}{cc}
0 & \frac{\theta (-n-\frac 12)t_1^n}{t_2^{n+1}} \\ 
\frac{\theta (n+\frac 12)t_2^n}{t_1^{n+1}} & 0
\end{array}
\right] -\theta (t_2-t_1)\left[ 
\begin{array}{cc}
0 & \frac{\theta (n+\frac 12)t_1^n}{t_2^{n+1}} \\ 
\frac{\theta (-n-\frac 12)t_2^n}{t_1^{n+1}} & 0
\end{array}
\right]  \label{w11}
\end{equation}
for $m=0.$ The function $F$ depends on eight anticommuting variables and is
therefore a polynomial of degree at most eight. It is straightforward to
check that for $t_1\neq 0$, the limits 
\begin{equation}
\lim_{t\rightarrow 0^{+}}\left. \tfrac \delta {t_1\cdot \delta \bar{\eta}%
_{-n}^{\uparrow }(t_1)}Z_n\tfrac{\overleftarrow{\delta }}{t\cdot \delta \eta
_n^{\uparrow }(t)}\right| _{\eta =\bar{\eta}=0}  \label{nr6}
\end{equation}
\begin{equation}
\text{and}\quad \lim_{t\rightarrow \infty }\left. \tfrac \delta {t_1\cdot
\delta \bar{\eta}_{-n}^{\uparrow }(t_1)}Z_n\tfrac{\overleftarrow{\delta }}{%
t\cdot \delta \eta _n^{\uparrow }(t)}\right| _{\eta =\bar{\eta}=0}
\label{r3z}
\end{equation}
are well-defined. Corresponding limits for the other two-point correlators
are also well-defined, and the same holds true for other correlators such as 
\begin{equation}
\lim_{t\rightarrow 0^{+}}\left. \tfrac \delta {t_1\cdot \delta \bar{\eta}%
_{-n-1}^{\downarrow }(t_1)}\tfrac \delta {t_3\cdot \delta \bar{\eta}%
_{-n}^{\uparrow }(t_3)}Z_n\tfrac{\overleftarrow{\delta }}{t_2\cdot \delta
\eta _{n+1}^{\downarrow }(t_2)}\tfrac{\overleftarrow{\delta }}{t\cdot \delta
\eta _n^{\uparrow }(t)}\right| _{\eta =\bar{\eta}=0}  \label{yy}
\end{equation}
\begin{equation}
\quad \text{and}\quad \lim_{t\rightarrow \infty }\left. \tfrac \delta
{t_1\cdot \delta \bar{\eta}_{-n-1}^{\downarrow }(t_1)}\tfrac \delta
{t_3\cdot \delta \bar{\eta}_{-n}^{\uparrow }(t_3)}Z_n\tfrac{\overleftarrow{%
\delta }}{t_2\cdot \delta \eta _{n+1}^{\downarrow }(t_2)}\tfrac{%
\overleftarrow{\delta }}{t\cdot \delta \eta _n^{\uparrow }(t)}\right| _{\eta
=\bar{\eta}=0}  \label{34x}
\end{equation}
provided that $t_1,t_2,t_3\neq 0.$ From this and the fact that $W_n$
vanishes when $\eta =\bar{\eta}=0$, we conclude that $F=0$ and we obtain (%
\ref{g4a}).

\TeXButton{vspace}{\vspace{12pt}}

\noindent \TeXButton{TeX field}
{\vspace{12pt}
\Large\bf{Appendix 2}
\normalsize\rm}

The following is a list of the spherical correlation functions. For $m\neq 0$
we have\footnote{%
See comments on Euclidean correlation functions immediately following (\ref
{aa6}).}

\begin{eqnarray}
\left\langle 0\left| \psi _n^{\uparrow }(t_1)\bar{\psi}_{-n}^{\uparrow
}(t_2)\right| 0\right\rangle _E &=&\theta (t_1-t_2)mK_n(\left| m\right|
t_1)I_n(\left| m\right| t_2)  \label{ra1} \\
&&+\theta (t_2-t_1)mI_n(\left| m\right| t_1)K_n(\left| m\right| t_2), 
\nonumber
\end{eqnarray}
\begin{eqnarray}
\left\langle 0\left| \psi _{n+1}^{\downarrow }(t_1)\bar{\psi}%
_{-n-1}^{\downarrow }(t_2)\right| 0\right\rangle _E\; &=&\theta
(t_1-t_2)mK_{n+1}(\left| m\right| t_1)I_{n+1}(\left| m\right| t_2)
\label{ra2} \\
&&+\theta (t_2-t_1)mI_{n+1}(\left| m\right| t_1)K_{n+1}(\left| m\right| t_2).
\nonumber
\end{eqnarray}
\begin{eqnarray}
\left\langle 0\left| \psi _n^{\uparrow }(t_1)\bar{\psi}_{-n-1}^{\downarrow
}(t_2)\right| 0\right\rangle _E &=&\theta (t_1-t_2)\left| m\right|
K_n(\left| m\right| t_1)I_{n+1}(\left| m\right| t_2)  \label{ra3} \\
&&-\theta (t_2-t_1)\left| m\right| I_n(\left| m\right| t_1)K_{n+1}(\left|
m\right| t_2),  \nonumber
\end{eqnarray}
\begin{eqnarray}
\left\langle 0\left| \psi _{n+1}^{\downarrow }(t_1)\bar{\psi}_{-n}^{\uparrow
}(t_2)\right| 0\right\rangle _E &=&\theta (t_1-t_2)\left| m\right|
K_{n+1}(\left| m\right| t_1)I_n(\left| m\right| t_2)  \label{ra4} \\
&&-\theta (t_2-t_1)\left| m\right| I_{n+1}(\left| m\right| t_1)K_n(\left|
m\right| t_2).  \nonumber
\end{eqnarray}
For $m=0$,

\begin{equation}
\left\langle 0\left| \psi _n^{\uparrow }(t_1)\bar{\psi}_{-n-1}^{\downarrow
}(t_2)\right| 0\right\rangle _E=\left[ 
\begin{array}{c}
\theta (t_1-t_2)\theta (-n-\tfrac 12) \\ 
-\theta (t_2-t_1)\theta (n+\tfrac 12)
\end{array}
\right] \tfrac{t_1^n}{t_2^{n+1}},  \label{h3}
\end{equation}
\begin{equation}
\left\langle 0\left| \psi _{n+1}^{\downarrow }(t_1)\bar{\psi}_{-n}^{\uparrow
}(t_2)\right| 0\right\rangle _E=\left[ 
\begin{array}{c}
\theta (t_1-t_2)\theta (n+\tfrac 12) \\ 
-\theta (t_2-t_1)\theta (-n-\tfrac 12)
\end{array}
\right] \tfrac{t_2^n}{t_1^{n+1}},  \label{5ex}
\end{equation}
\begin{equation}
\left\langle 0\left| \psi _n^{\uparrow }(t_1)\bar{\psi}_{-n}^{\uparrow
}(t_2)\right| 0\right\rangle _E=\left\langle 0\left| \psi _{n+1}^{\downarrow
}(t_1)\bar{\psi}_{-n-1}^{\downarrow }(t_2)\right| 0\right\rangle _E=0.
\label{ve33}
\end{equation}

\end{document}